# Estabelecimento do Conceito de Temperatura como uma grandeza derivada da Energia e da Entropia.

Rodrigo de Abreu

Departamento de Física do IST, Universidade Técnica de Lisboa

Abstract

Temperature is introduced as a derived concept from energy and entropy. We consider two sub-systems in equilibrium for several configurations. The equality of temperature of the sub-systems is obtained from the equilibrium condition. The isothermal transformation is defined for several configurations and from this definition we obtain the Clausius-Clapeyron equation. We apply the analysis to the ideal gas. The classical ideal gas appears as a limit and the problem of the measurement of temperature is analysed.

Resumo

O conceito de equilíbrio termodinâmico pode ser conceptualizado através da estacionaridade da grandeza entropia: uma transformação em que a entropia não varia (isentrópica) é uma sucessão de estados de equilíbrio termodinâmico. Mostra-se que ao longo duma isentrópica de um Sistema de energia $U$ entropia $S$ e volume $V$, constítuido por dois sub-sistemas de energia $U_i$, entropia $S_i$ e volume $V_i$, a temperatura, definida como $\lambda=(\partial U/\partial S)_V$, é igual à temperatura dos sub-sistemas, $\lambda_i=(\partial U_i/\partial S_i)_{V_i}$ ($i=1,2$), e é constante ao longo de uma transformação em que $(\partial p_1/\partial T)_{V_1}/(\partial p_2/\partial T)_{V_2}=-dV_2/dV_1$, em que $T=\lambda$ e $p_i$ é a pressão do sub-sistema $i$. Obtem-se a equação de Clausius-Clapeyron e o limite superior do rendimento de um motor.
Da aplicação da teoria ao gás ideal resulta a calibração de um termómetro.

## Introdução

Um estado de equilíbrio termodinâmico caracteriza-se pela constância no tempo dos diversos parâmetros que podem ser concebidos e medidos no sistema. Num estado de equilíbrio temos um dado volume, uma dada energia e uma dada entropia [1, 2]. Se alterarmos as ligações a que o sistema está sujeito, o sistema (conjunto de susb-sistemas) aumenta de entropia até que se atinja um novo estado de equilíbrio compatível com as novas ligações.

O sistema que se vai considerar é constituído por dois sub-sistemas de energia $U_i$, volume $V_i$ e entropia $S_i$ ($i=1, 2$). Os sub-sistemas trocam energia através duma parede fixa e cada um deles pode variar de volume através do movimento de um êmbolo. Este movimento deve-se à acção de uma força exterior. O trabalho destas forças (cada sub-sistema tem um êmbolo e cada êmbolo é submetido a uma força) altera eventualmente a energia do sistema. Se o trabalho de uma das forças for compensado pelo trabalho da outra força que actua no outro sub-sistema a energia do sistema permanece constante. No entanto a energia dos sub-sistemas pode variar, mesmo quando a energia do Sistema não varia.

Num estado de equilíbrio a entropia do sistema S tem um determinado valor. Numa transformação em que as forças exteriores estão em equilíbrio com as forças interiores [1, 2] a entropia do sistema não varia embora dado se ter imposto $S=S_1 + S_2$, $dS_1$ e $dS_2$ sejam simétricos. Se a transformação não for reversível a entropia aumenta.

Em **I.** mostra-se que em pontos de equilíbrio o parâmetro $\lambda=(\partial U/\partial S)_V$ é igual a $\lambda_i =(\partial U_i/\partial S_i)_{Vi}$. Introduz-se a temperatura $T$ identificando-a com o parâmetro $\lambda=(\partial U/\partial S)_V$.

Em **II.** mostra-se que agindo reversívelmente sobre o sistema, variando os volumes de quantidades infinitesimais $dV_1$ e $dV_2$ de tal forma que $dV_2/dV_1$ seja simétrico das relações entre as derivadas parciais das pressões em ordem à temperatura

$(\partial p_1/\partial T)_{V1}/(\partial p_2/\partial T)_{V2}$

a temperatura não varia, para os diversos pares de volumes ($V_1$, $V_2$) que vão sendo definidos. Para o gás ideal clássico e para o gás de fotões esta condição pode ser posta na forma da relação entre pressões o que significa, neste caso, que a energia do sistema, conjunto de sub-sistemas, também não varia.

Desta forma deriva-se a partir da energia e da entropia a condição de igualdade de temperaturas de sub-sistemas em equilíbrio termodinâmico e a condição de definição de uma transformação isotérmica.

Em **III.** aplica-se a teoria ao caso particular de um gás ideal, gás de partículas sem interacção recíproca. O gás ideal clássico surge como caso limite.

Em **IV.** obtem.se a equação de Clausius-Clapeyron.

Em **V.** calibra-se um termómetro. As hipóteses que estão associadas a esta calibração surgem de forma simples e clara.

Em **VI.** determina-se o rendimento máximo de um motor.

# I. A igualdade das temperaturas de dois sub-sistemas em equilíbrio termodinâmico.

Consideremos dois sub-sistemas em interacção de acordo com a fig. 1

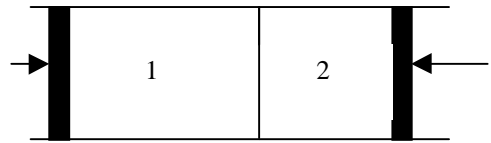

*Fig. 1*

**Sistema constituído por dois sub-sistemas 1 e 2. Interagem entre si através de uma parede fixa e com o exterior através de dois êmbolos.**

A energia do sistema constituído pelos sub-sistemas 1 e 2 é

$$U = U_1 + U_2, \qquad (1)$$

e a entropia é

$$S = S_1 + S_2. \qquad (2)$$

Desprezamos o termo de interacção no valor absoluto de $U$ e de S.

Dado que $U_i = U_i(V_i, S_i)$ temos

$$dU_i = (\partial U_i/\partial V_i)_{S_i} dV_i + (\partial U_i/\partial S_i)_{V_i} dS_i. \qquad (3)$$

Como para $dS_i = 0$, $dU_i = -p_i dV_i$, temos de (1) e (3)

$$dU = -p_1\,dV_1 + \lambda_1\,dS_1 - p_2\,dV_2 + \lambda_2\,dS_2, \qquad (4)$$

em que

$$\lambda_i = (\partial U_i/\partial S_i)_{Vi}, \ (i=1,\ 2). \qquad (5)$$

Admitamos que o sistema se encontra num estado de equilíbrio. Uma perturbação em torno deste estado permite escrever de (1) e considerando que $U = U(V_1,\ V_2,\ S)$,

$$dU = (\partial U_1/\partial V_1)_{S,V1}\,dV_1 + (\partial U_2/\partial S_2)_{V2}\,dV_2 + \lambda\,dS \qquad (6)$$

em que

$$\lambda = (\partial U/\partial S)_V. \qquad (7)$$

Como numa transformação reversível ($dS = 0$)

$$dU = -p_1\,dV_1 - p_2\,dV_2,$$

temos de (6)

$$p_i = -(\partial U/\partial V_i)_{S,Vj},\ (i = 1,\ 2;\ j = 2,\ 1). \qquad (8)$$

Deste modo (6) escreve-se

$$dU = -p_1\,dV_1 - p_2\,dV_2 + \lambda\,dS \qquad (9)$$

Consideremos dS=0. De (9) e (4) temos que

$$\lambda_1\,dS_1 + \lambda_2\,dS_2 = 0 \qquad (10)$$

e como $dS = dS_1 + dS_2$, $dS_1 = -dS_2$,

$$\lambda_1 = \lambda_2. \qquad (11)$$

Mas de (9) e (4) temos,

$$\lambda\, dS = \lambda_1\, dS_1 + \lambda_2\, dS_2, \tag{12}$$

e verificando-se (11)

$$\lambda\, dS = \lambda_1\, (dS_1 + dS_2) = \lambda_2\, (dS_1 + dS_2), \tag{13}$$

e portanto temos que

$$\lambda = \lambda_1 = \lambda_2. \tag{14}$$

Em pontos de equilíbrio, verifica-se a igualdade das temperaturas entre os sub-sistemas, e qualquer destas temperaturas é a temperatura do sistema, $T = \lambda = (\partial U/\partial S)_V$.

## II. Definição da transformação isotérmica.

Consideremos que $S=S(V,T)$. Temos diferenciando

$$dS = (\partial S /\partial T)\, dT + (\partial S /\partial V)\, dV. \tag{15}$$

De uma das relações de Maxwell temos

$$(\partial S /\partial V) = (\partial p /\partial T), \tag{16}$$

e portanto de (15)

$$dS = (\partial S /\partial T)\, dT + (\partial p /\partial T)\, dV. \tag{17}$$

Se $dT=0$, temos de (17)

$$dS = (\partial p /\partial T)\, dV. \tag{18}$$

Esta relação é válida para cada um dos sub-sistemas i e portanto ao longo de pontos de equilíbrio em que a entropia e a temperatura não variam temos

$$dS = dS_1 + dS_2 = (\partial p_1/\partial T)\, dV_1 + (\partial p_2/\partial T)\, dV_2. \quad (19)$$

Os volumes vão variando de acordo com a relação

$$(\partial p_1/\partial T)/(\partial p_2/\partial T) = -dV_2/dV_1. \quad (20)$$

Consideremos a fig. 2:

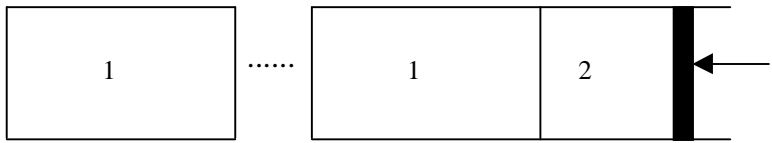

*Fig. 2*

*Para uma dada configuração da fig. 1, podemos conceber a fig. 2 em que se justapõe a 2 N cópias de 1. Se N for muito grande, variando o volume de 2 a energia que passa para as N cópias do sub-sistema 1 apenas altera a energia de cada uma das cópias de uma quantidade fisicamente pequena. Deste modo a temperatura do sub-sistema 2, em pontos de equilíbrio é a temperatura de cada uma das cópias que por sua vez é a temperatura das N cópias, a temperatura da "fonte de calor" constituída pelas N cópias de 1.*

A transformação isotérmica está deste modo definida.

## III. Aplicação a um gás ideal.

Um gás de partículas que não interagem entre si satisfaz, em determinadas condições [3, 4, 5, 6, 7, 8, 9, 10, 11], à equação $p = \alpha u$ em que $u$ é a densidade de energia, $p$ a pressão e $\alpha$ uma constante cujo valor depende do gás que se considera. Exemplifiquemos com um gás ideal monoatómico em que $\alpha = 2/3$ e com um gás de fotões em que $\alpha = 1/3$. Dado que qualquer gás está sempre na presença de um gás de fotões, só como aproximação é que podemos pensar num gás sem fotões (por esta razão *Humpry Davy* referia-se ao oxigênio como *phoxigénio*).

De $p = \alpha\, u$ pode fácilmente obter-se a relação $pV = B\,\lambda$ em que $B$ só é constante ao longo de uma isentrópica, isto é $B = B(\lambda V^\alpha)$ dado $\lambda V^\alpha$ ser constante ao longo de uma isentrópica. Como $pV = \alpha\, U$ temos que $\alpha U = B\,\lambda$, isto é $U = A\,\lambda$ em que $A = B/\alpha$. De facto:

Como $p = -(\partial U/\partial V)_S$ temos que

$$(\partial p/\partial S)_V = -\partial^2 U/(\partial S\,\partial V). \qquad (21)$$

Como $\lambda = (\partial U/\partial S)_V$ temos

$$(\partial \lambda/\partial V)_S = -\partial^2 U/(\partial V\,\partial S), \qquad (22)$$

ou seja

$$(\partial p/\partial S)_V = -(\partial \lambda/\partial V)_S. \qquad (23)$$

Se $p = \alpha\, U/V$, temos

$$(\partial p/\partial S)_V = \alpha\,(\partial U/\partial S)/V = \alpha\,\lambda/V. \qquad (24)$$

De (23) e (24) vem

$$(\partial \lambda/\partial V)_S = -\alpha\,\lambda/V, \qquad (25)$$

ou, integrando

$$\lambda V^\alpha = c^{te}., \qquad (26)$$

ao longo de uma isentrópica.

De $p = \alpha u$ e $p = -(\partial U/\partial V)_S$ temos

$$(\partial U/\partial V)_S = -\alpha\, U/V, \qquad (27)$$

que, integrando, dá origem à relação

$$UV^\alpha = c^{te}., \qquad (28)$$

ao longo de uma isentrópica. Comparando (28) e (26) temos

$$U = A\,\lambda \qquad (29)$$

em que A é constante ao longo de uma isentrópica, isto é A é uma função de $\lambda V^\alpha$, $A = A(\lambda V^\alpha)$. Dado $p\,V = \alpha\,U$, temos de (29)

$$p\,V = \alpha\,A\,\lambda = B\lambda = BT. \qquad (30)$$

Define-se um gás ideal clássico pela condição $A = c^{te}.$. Para um gás ideal clássico a energia não varia ao longo de uma isotérmica.

Admitamos que os sub-sistema 1 e 2, considerados anteriormente em **I.** são gases ideais clássicos. Ao longo de uma isotérmica como para o gás ideal clássico $U = c^{te}$ e dado que $S = c^{te}$ temos de (6)

$$dU = 0 = -p_1\,dV_1 - p_2\,dV_2\,, \qquad (31)$$

isto é

$$dV_1 = -(p_2/p_1)\,dV_2. \qquad (32)$$

De (30), (32) permite escrever

$$dV_1 = -(B_2\lambda_2/B_1\lambda_1)(V_1/V_2)\,dV_2, \qquad (33)$$

ou

$$dV_1 / V_1 = -(B_2 / B_1) \, dV_2 / V_2 \qquad (34)$$

ou, integrando

$$\ln V_1 = -\ln V_2^{(B2/B1)}, \qquad (35)$$

ou

$$V_1^{B1} V_2^{B2} = c^{te}.. \qquad (36)$$

A relação (36) define os pares ($V_1$, $V_2$) tais que a temperatura não varia, em que $U_1$ e $U_2$ não variam.

Analisemos seguidamente a isotérmica para um gás de fotões:

A densidade de energia dos fotões é constante ao longo da isotérmica, $u = u(\lambda)$. È como se não existisse a parede que divide os sub-sistemas 1 e 2 ou é como se existisse uma janela transparente, que permite a passagem dos fotões; isto é, $u_1 = u_2$ e portanto $u = u(\lambda)$. Para os fotões $\alpha = 1/3$ e portanto (28) escreve-se

$$UV^{1/3} = c^{te}.,$$

ou

$$U^3 V = c^{te}., \qquad (37)$$

ou, ainda

$$U^4 V/U = c^{te}, \qquad (38)$$

$$U/V = b\ U^4 \qquad (39)$$

em que *b* é constante ao longo de uma isentrópica. Como ao longo de uma isentrópica $U = A\ \lambda$ temos, de (39)

$$u = b\ A\ \lambda^4 = a\ \lambda^4 = a\ T^4 \qquad (40)$$

em que *a* é constante ao longo de uma isentrópica. Mas como ao longo de uma isotérmica *u* é uma constante, dado $u = u(\lambda)$, concluímos que

$$u = a\ \lambda^4 = a\ T^4 \qquad (41)$$

é uma relação universal com $a = c^{te}$..

Temos, portanto, que para os fotões também se verifica (31), tendo em atenção que $p = \alpha\ u = \alpha\ a\ \lambda^4$ e portanto $p_1 = p_2 = p$ ao longo de uma isotérmica

$$-p\ dV_1 - p\ dV_2 = 0 \qquad (42)$$

ou

$$dV_1 = -\ dV_2,\ \text{isto é}$$

$$V_1 + V_2 = c^{te}.. \qquad (43)$$

Consideremos finalmente um gás ideal constituído por um gás de fotões e por um gás ideal em que *B* varie muito lentamente.

Para um dos sub-sistemas *i* temos

$$P_i = B_i \lambda\ / V_i + a/3\ \lambda^4. \qquad (44)$$

De (20) temos

$$dV_1 = -(B_2/V_2 + a/3\ 4\lambda^3/(B_1/V_1 + a/3\ 4\lambda^3))\ dV_2 \quad (45)$$

ou

$$(B_1/V_1)\ dV_1 + a/3\ 4\lambda^3\ dV_1 = -(B_2/V_2)\ dV_2 - a/3\ 4\lambda^3\ dV_2. \quad (46)$$

Integrando, vem

$$B_1\ lnV_1 + a/3\ 4\lambda^3\ V_1 = -B_2\ lnV_2 - a/3\ 4\lambda^3\ V_2 + c^{te}., \quad (47)$$

ou

$$\lambda\ ln(V_1^{B_1}\ V_2^{B_2}) + a/3\ 4\lambda^4\ (V_1 + V_2) = c^{te}.. \quad (48)$$

(48) é a isotérmica para a "mistura" dos dois gases.

O gás ideal clássico surge assimptóticamente no limite quando a temperatura tende para zero, em que se anula completamente o efeito do radiamento.

## IV. A equação de Clausius-Clapeyron.

Se num dos sub-sistemas se encontrarem duas fases, por exemplo liquido-vapor, a equação (20) permite obter a equação de Clausius-Clapeyron. De facto temos de (20) e admitindo que no sub-sistema (1) se encontram as duas fases, temos

$$(\partial p_1/\partial T) = -(\partial p_2/\partial T)\ dV_2/dV_1 \quad (49)$$

e de (3) temos para $i=2$

$$dU_2 = (\partial U_2/\partial V_2)_{S_2}\ dV_2 + (\partial U_2/\partial S_2)_{V_2}\ dS_2, \quad (50)$$

e dado que na mudança de estado $dT=0$

$$dU_2 = -p_2 dV_2 + TdS_2 = -p_2 dV_2 + T(\partial S_2/\partial V_2)_T\, dV_2 \qquad (51)$$

ou seja

$$dU_2 = -p_2 dV_2 + T(\partial p_2/\partial T)_{V_2}\, dV_2, \qquad (52)$$

ou

$$(\partial p_2/\partial T)_{V_2}\, dV_2 = (dU_2 + p_2 dV_2)/T = dQ/T. \qquad (53)$$

Integrando entre dois volumes do sub-sistema (1), $V_{1i}$ e $V_{1f}$ obtemos a equação de Clausisus-Clapeyron

$$(\partial p_1/\partial T) = L/(T(V_{1f} - V_{1i})) \qquad (54)$$

em que $L$ é o chamado calor latente de mudança de estado, a energia que passa através da parede que separa os sub-sistemas (1) e (2), na mudança de estado correspondente aos volumes $V_{1i}$ e $V_{1f}$.

## V. A calibração de um termómetro.

Se desprezarmos a pressão do radiamento o volume de um gás ideal é dado por

$$V = (B/p)\,\lambda. \qquad (55)$$

O coeficiente de expansão térmica é definido por

$$\alpha = (1/V)(\partial V/\partial \lambda)_p \qquad (56)$$

De (55) temos

$$\alpha = B/(pV) = B/B\lambda = 1/\lambda \qquad (57)$$

A medição de $\alpha$ é a medição de $\lambda$. Para medirmos $\alpha$ necessitamos de determinar $(\partial V/\partial \lambda)_p$. Admitir uma relação linear entre $V$ e $\lambda$, é admitir um valor constante para $B$. É de admitir que *B* tenda assimptóticamente para um determinado valor quando a pressão tende para zero. A experiência mostra que gases diferentes a baixas pressões ocupando o mesmo volume quando à mesma pressão e á mesma temperatura, sofrem aproximadamente a mesma variação de volume quando sofrem a mesma variação de temperatura e que esta variação é tanto mais aproximada quanto mais baixa for a pressão. Por outro lado, para baixas pressões verifica-se que para um mesmo valor de $\lambda$, o produto da pressão pelo volume é também aproximadamente constante, e o valor desta constante é independente do gás e independente de $\lambda$. Estes resultados permitem, por extrapolação, determinar o valor de *B* quando a pressão tende para zero. Para tal consideram-se duas temperaturas, por exemplo o ponto de fusão do gelo e o ponto de ebulição da água à pressão de uma atmosfera, e considera-se que existem entre estas duas temperaturas n intervalos unitários $\Delta \lambda$. Este *n*, evidentemente, é arbitrário. Determina-se o valor de $\Delta V$ e divide-se por *n*. Para pressões a tender para zero determina-se por extrapolação o valor de *B*, através dos sucessivos valores de $(\partial V/\partial \lambda)_p$ = *B/p*, para sucessivos valores decrescentes de *p*. Para uma dada temperatura, determina-se por extrapolação $\alpha = B/(pV)$, dado para sucessivos valores decrescentes de *p*, $\alpha = B/(pV)$ vai-se se aproximando de $B_0/[lim_{p->0}(pV)]$, dado o $[lim_{p->0}(pV)]$ ser uma constante e que se pode determinar experimentalmente para os sucessivos valores decrescentes da pressão. $\lambda$ é $1/\alpha$. À escala de temperaturas em que se arbitrou para *n* o valor 100 chamou-se escala *Kelvin* e a temperatura simbolizou-se por *T*. Em 1954 na *General Conference of Weights and Measures* atribuiu-se em face dos resultados experimentais de então, o valor de 273,15 *K* para a temperatura de fusão do gelo à pressão de 1 *atm* [12]. Resultados recentes levaram a que se tenha redefinido o valor de *n* que já não é rigorosamente 100, para que o valor do ponto de fusão do gelo permaneça 273,15 *K*. É evidente, pelo anteriormente afirmado, que a calibração de um termómetro é matéria delicada, em contínuo aperfeiçoamento e sofisticação, dado que não basta calibrar o termómetro apenas em alguns pontos fixos. É necessário calibrar o termómetro em pontos intermédios e fora do intervalo definido pelos pontos fixos [12]. Para temperaturas elevadas a calibração dos termómetros é feita usando o gás de fotões.

## VI. O rendimento de um motor e a temperatura Kelvin.

Consideremos duas fontes de calor (ver fig. 2), cujas temperaturas *Kelvin* são $T_2$ e $T_1$. Consideremos uma substância que descreve um ciclo trocando energia com duas fontes, regressando, portanto, ao estado inicial. A variação de entropia global neste ciclo é maior ou igual a zero. Para o ciclo, o rendimento é

$$\eta = |W|/|\Delta U_{F2}| \qquad (58)$$

em que $W$ e $\Delta U_{F2}$ são respectivamente o trabalho realizado pelas forças exteriores e a variação de energia da fonte quente, e, estamos a impor, $W$ é negativo ou nulo (motor) e $\Delta U_{F2} \leq 0$ (fonte quente).

Dado que a variação de entropia global é maior ou igual a zero e dado que a substância que constitui o motor regressa ao estado inicial (tem portanto no ciclo uma variação de entropia nula), temos

$$\Delta S_{F1} + \Delta S_{F2} \geq 0, \qquad (59)$$

ou seja a relação (3), aplicada à fonte, cujo volume é constante, permite escrever de (59)

$$\Delta U_{F1}/T_1 + \Delta U_{F2}/T_2 \geq 0. \qquad (60)$$

Do princípio de conservação de energia temos quando o motor descreve um ciclo

$$W = \Delta U_{F2} + \Delta U_{F1}. \qquad (61)$$

Como $W \leq 0$, tem-se de (61) $|\Delta U_{F2}| \geq |\Delta U_{F1}|$

De (59), tendo em atenção que a variação de energia da fonte quente é negativa ou nula, e que a variação de energia da fonte fria é positiva ou nula,

$$-|\Delta U_{F2}|/T_2 + |\Delta U_{F1}|/T_1 \geq 0, \qquad (62)$$

ou

$$|\Delta U_{F1}|/T_1 \geq |\Delta U_{F2}|/T_2, \qquad (63)$$

ou

$$|\Delta U_{F1}|/|\Delta U_{F2}| \geq T_1/T_2. \qquad (64)$$

De (58), (61) e (64) temos

$$\eta = 1 - |\Delta U_{F1}|/|\Delta U_{F2}| \qquad (65)$$

$$\eta \leq 1 - T_1/T_2. \qquad (66)$$

Conclusão

Mostrou-se que o conceito de temperatura pode ser estabelecido como grandeza derivada da energia e da entropia. Para tal considerou-se um sistema constituído por dois sub-sistemas em equilíbrio termodinâmico. Em pontos de equilíbrio a temperatura é definida como a variação da energia com a entropia quando não variam as variáveis de deformação, no caso considerado o volume. A transformação isotérmica, a transformação em que a temperatura não varia, é a transformação em que as variações elementares dos volumes dos sub-sistemas é simétrica da relação inversa das derivadas parciais das pressões dos sub-sistemas em ordem à temperatura. Para o gás ideal clássico e para o gás de fotões, na isotérmica, a energia e a entropia do sistema, conjunto de sub-sistemas, não variam. Obteve-se desta definição de isotérmica a equação de Clausius-Clapeyron. A partir da conceptualização de um gás ideal mostrou-se como é possivel calibrar um termómetro. Obteve-se de forma simples e directa o rendimento máximo dos motores térmicos.